\documentclass[final,3p,times,twocolumn,longtitle]{elsarticle}

\journal{Nuclear Instrument and Method}

\usepackage{graphicx}
\usepackage{lineno}
\usepackage{color}

\begin{document}

\begin{frontmatter}

\title{A liquid scintillator for a neutrino Detector working at -50 degree}

\author[ihep,ucas]{Zhangquan Xie\protect\footnote{Email: xiezq@ihep.ac.cn.}}
\author[ihep,ucas]{Jun Cao\protect\footnote{Email: caoj@ihep.ac.cn.}}
\author[ihep]{Yayun Ding}
\author[ihep]{Mengchao Liu}
\author[ihepklab]{Xilei Sun}
\author[ihep,NJU]{Wei Wang}
\author[ihepklab]{Yuguang Xie}

\address[ihep]{Institute of High Energy Physics, Chinese Academy of Sciences, Beijing 100049, China}
\address[ihepklab]{State Key Laboratory of Particle Detection and Electronics, Institute of High Energy Physics, Chinese Academy of Sciences, Beijing 100049, China}
\address[ucas]{University of Chinese Academy of Sciences, Beijing 100049, China}
\address[NJU]{Nanjing University, Nanjing, 210093, China}
\begin{abstract}
\noindent A liquid scintillator (LS) is developed for the Taishan Antineutrino Observatory (TAO), a ton-level neutrino detector to measure the reactor antineutrino spectrum with sub-percent energy resolution by adopting Silicon Photomultipliers (SiPMs) as photosensor. To reduce the dark noise of SiPMs to an acceptable level, the LS has to work at -50$^\circ$C or lower. A customized apparatus based on a charge-coupled device (CCD) is developed to study the transparency of the liquid samples in a cryostat. We find that the water content in LS results in transparency degradation at low temperature, which can be cured by bubbling dry nitrogen to remove water. Adding 0.05\% ethanol as co-solvent cures the solubility decrease problem of the fluors PPO and bis-MSB at low temperature. Finally, a Gadolinium-doped liquid scintillator (GdLS), with 0.1\% Gd by weight, 2~g/L PPO, 1~mg/L bis-MSB, and 0.05\% ethanol by weight in the solvent LAB, shows good transparency at -50$^\circ$C and also good light yield.
\end{abstract}

\begin{keyword}
Scintillation detectors \sep Neutrino oscillations \sep Reactor
\PACS 29.40.Mc \sep 14.60.Pq \sep 28.41.-i
\end{keyword}

\end{frontmatter}


\section{Introduction}

Liquid scintillator (LS) has been used to detect reactor antineutrinos since 1950s~\cite{Cowan1956}. The hydrogen-rich LS serves as both the electron antineutrino target and the active detection material. In the classic detection channel, a reactor antineutrino produces a positron and a neutron via Inverse Beta decay (IBD) reaction, $\bar\nu_e \ + \ p \rightarrow e^+ \ + \ n $. The positron energy measured with scintillation light is an excellent proxy for the antineutrino energy. The best energy resolutions achieved by large-scale LS detectors are $6.4\%/\sqrt{E(\mathrm{MeV})}$ ($\sim300$ p.e./MeV) for the 1-kton KamLAND detector~\cite{kamland2011} and $5\%/\sqrt{E(\mathrm{MeV})}$ ($\sim500$ p.e./MeV) for the 500-ton Borexino detector~\cite{borexino2008}, where p.e.\ stands for photoelectrons. The 20-kton JUNO detector is designed with an energy resolution of $3\%/\sqrt{E(\mathrm{MeV})}$ ($\sim1200$ p.e./MeV)~\cite{An:2015jdp}, almost the limit of conventional technology with photomultipliers (PMTs).

Recently the shape of the reactor antineutrino spectrum measured by Daya Bay~\cite{An:2015nua}, Double CHOOZ~\cite{Abe:2014bwa}, NEOS~\cite{Ko:2016owz}, and RENO~\cite{Bak:2018ydk} shows more than 10\% deviation from model predictions~\cite{Huber:2011wv, Mueller:2011nm, Estienne:2019ujo}. The Taishan Antineutrino Observatory (TAO)~\cite{junotaocdr} is thus proposed to measure the fine structure in the reactor antineutrino spectrum with as high as possible energy resolution. Such a measurement will provide a reference spectrum for JUNO, and also serve as a benchmark to test nuclear databases. TAO will adopt new technology to reach an energy resolution of $<2\%/\sqrt{E(\mathrm{MeV})}$ ($\sim4500$ p.e./MeV), by viewing the Gadolinium-doped LS (GdLS) with Silicon Photomultipliers (SiPMs) of photon detection efficiency $>50$\%.

The SiPMs have to operate at -50$^\circ$C or lower, to reduce their dark noise to an acceptable level for single p.e.\ detection. It might be possible to design a detector in which the SiPMs work at low temperature while the GdLS keeps at higher temperature. The LS light yield is a function of temperature. To have a high precision detector with good energy uniformity and stability, the temperature uniformity in the GdLS volume of TAO is required to be $<1$$^\circ$C. However, continuous heat flow from the GdLS to the SiPMs will make the thermal stability of the detector very difficult to maintain. Therefore, TAO will have the whole detector working at -50$^\circ$C, including the GdLS.

Many studies on properties of various LS at low temperature have been done. For examples, as early as in 1956, Ziegler {\it et al.}~\cite{Ziegler1956} reported an increase by a factor of 1.2 in the scintillation pulse height for LS with xylene as solvent, when lowering the temperature from 30$^\circ$C to -35$^\circ$C. 2,5-diphenyloxzzole (PPO) solutions of aromatic hydrocarbons, e.g. benzene, o-xylene, m-xylene, ethylbenzene and cumene were found to shift toward higher pulse heights with decreasing temperature~\cite{HOMMA198791}. The light yield of Linear Alkylbenzene (LAB) based LS was found to increase $\sim0.3$\% per degree in average from room temperature to -40$^\circ$C~\cite{Xia:2014cca}. A similar result was also reported in another study from 30$^\circ$C to -5$^\circ$C~\cite{Sorensen:2018skx}. However, all these studies focus on reducing thermal quenching, thus increasing the light yield of LS, with small samples. For a large-scale detector, where the light attenuation of LS is important, detailed studies are still lacking.

In this article, we report several potential problems found for the LS at low temperature, then develop a new recipe suitable for the ton-level TAO detector, based on the recipe for Daya Bay~\cite{Beriguete:2014gua} and for JUNO~\cite{Djurcic:2015vqa}. Since we find that doping Gd or not has little impact to the LS behavior at low temperatures, we will not distinguish LS and GdLS in this paper unless necessary. In the following, first we will introduce the experimental setup to study the light transmission at low temperature. Then we study the light transmission of the LS solvent and LS with different concentrations of fluors. We cure the solubility decrease problem by adding co-solvent into the LS. Light yield of the new LS is also discussed.

\section{Light transmission measurement apparatus}

To measure the transparency of the liquid scintillator at low temperature, the sample and the key components of the apparatus are placed in a cryostat of about 1 m$^3$ volume. In the beginning, a portable Ultraviolet–visible spectrometer (UV-vis) has been used to measure the sample in a 10-cm quartz cuvette. It is found that the measurements are not stable. When changing the temperature, the light spot of the transmitted light varies as the index of refraction of the liquid sample changes. The UV-vis could not provide a stable measurement for such cases.

\begin{figure}[htb]
\begin{center}
\includegraphics[width=8cm]{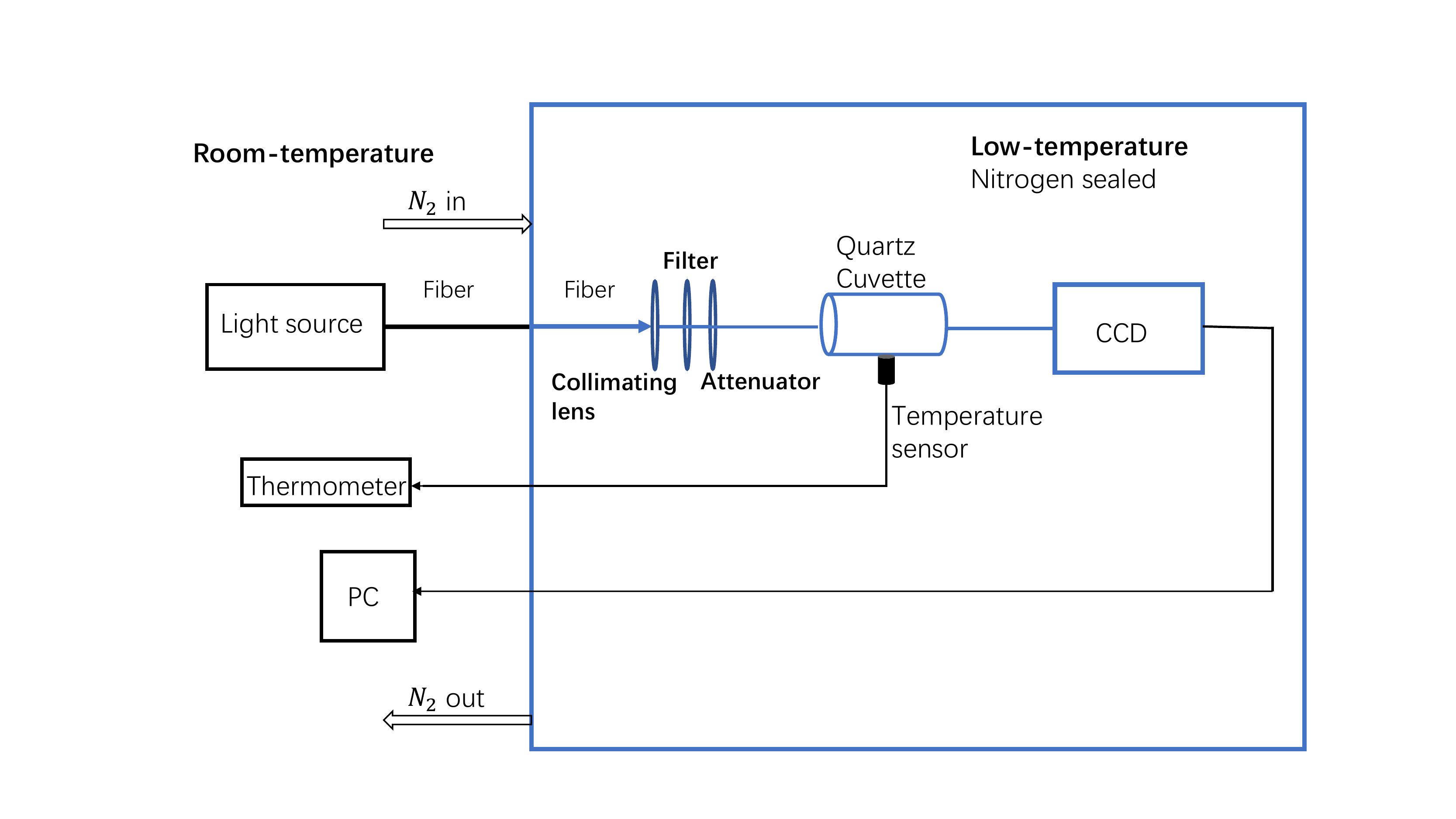}
\caption{Scheme of the apparatus measuring the light transmission of liquid samples at low temperature. The blue box represents the cryostat. \label{fig:appratus}}
\end{center}
\end{figure}

An experimental setup with a Charge Coupled Device (CCD) as the photosensor has been built to study the light transmission, the scheme of which is shown in Fig.~\ref{fig:appratus}. The light source is a DH-2000-BAL deuterium \& tungsten halogen lamp, covering the wavelength range from 200~nm to 900~nm. A temperature-insensitive optical fiber guides the light from the light source into the cryostat. After passing a collimating lens and a monochromatic filter at 430~nm, the light intensity is further reduced to 10\% with an attenuator, since the dynamic range of the CCD is $\leq$ 40,000 electrons/pixel. The liquid sample is filled in a cylindric quartz cuvette of 2 cm in diameter and 10 cm in length. The transmitted light spot and its intensity is detected by a KAF-16200 CCD of 4500 (H) $\times$ 3600 (V) pixels. A PT100 temperature sensor is pasted on the surface of the cuvette to monitor temperature of the sample. Water vapor in air should be very carefully prevented during the test. The whole cryostat is thus continuously flushed with dry nitrogen to prevent frost formation on the optical surfaces of the components and avoid absorption of water into the liquid sample during the test at low temperature.

The intensity of the transmitted light is obtained by integrating the charge of the CCD pixels. The dark noise is determined and subtracted for each measurement by comparing the charge with and without light input, which is at the level of 0.5\% of the signal at 20$^\circ$C and 0.2\% at -50$^\circ$C. In this study, we are interested in the relative transmittance at low temperatures comparing to that at room temperature, while the absolute attenuation length of the liquid, usually too long to be measured with a 10-cm cuvette, can be measured with other methods at room temperature. Another advantage of using a CCD is that we can compare the profiles of the light spots at low and room temperatures, to qualitatively understand the variations.

\begin{figure}[htb]
  \centering
  \includegraphics[width=7cm]{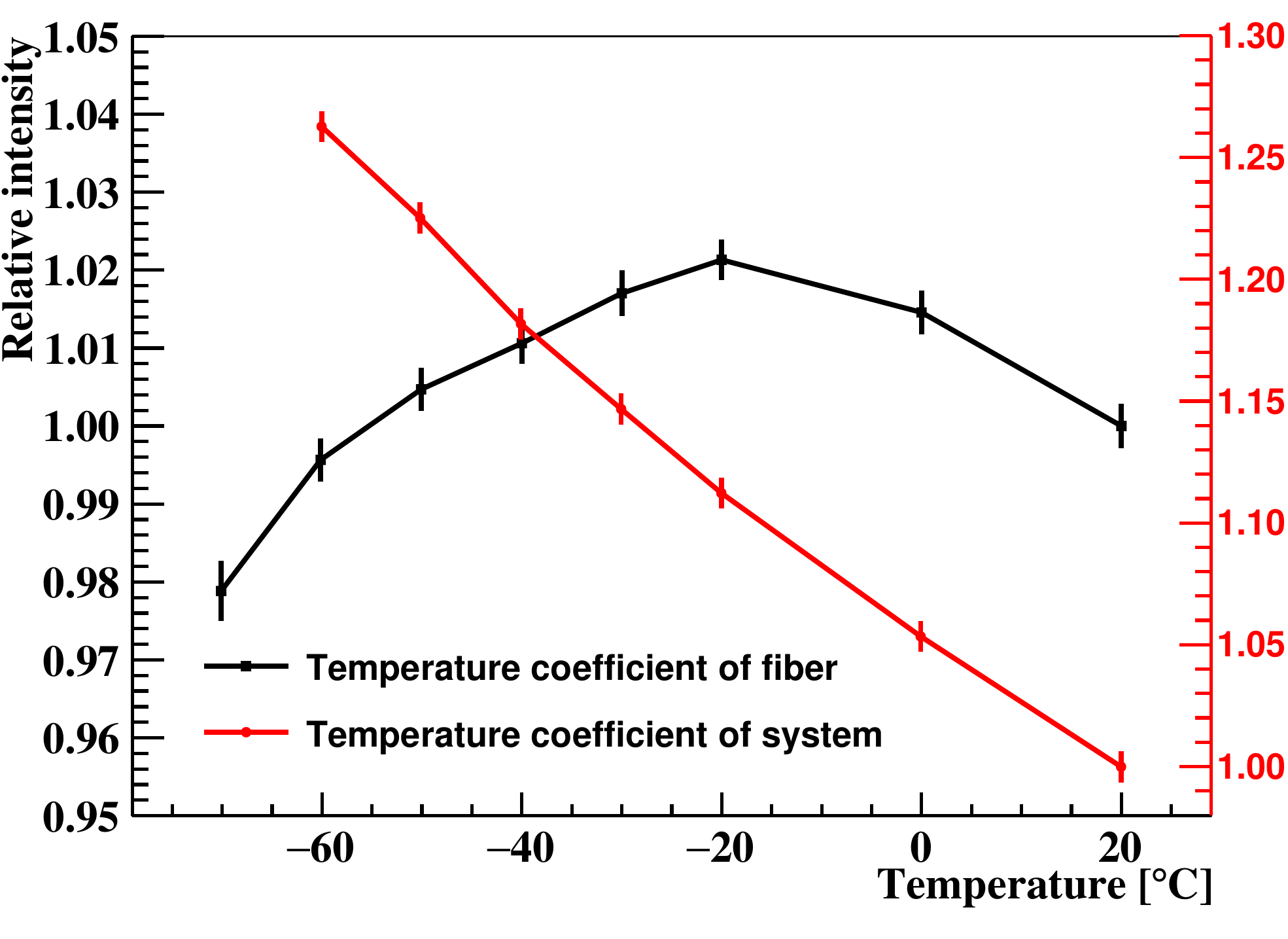}
\caption{The temperature dependence of the relative attenuation coefficient of the optical fibre is shown in black solid line. The combined temperature effect of the CCD and fibre is shown in red solid line.  Both are normalized to unity at 20$^\circ$C. \label{fig:appstability} }
\end{figure}

Stability of the light source has been studied by operating the apparatus at room temperature and monitoring the CCD response versus time. The light intensity drops by $\sim1.2$\% in the first 60 minutes and then keeps stable. With multiple measurements, we determine that the stability is within $0.15$\%. The temperature dependence of the relative attenuation coefficient of the optical fibre is measured to be in a range of 4\% between -50$^\circ$C and 20$^\circ$C as shown in Fig.~\ref{fig:appstability}, by moving the CCD out of the cryostat and viewing the light through a quartz window on the cryostat wall. Impacts of the refractive index change of quartz (the cryostat window and the cuvette) is negligible. The combined temperature effect of the CCD and the fibre is then measured with the scheme shown in Fig.~\ref{fig:appratus} with an empty cuvette. The relative light intensity increases from 1.0 at 20$^\circ$C (the normalization point) to 1.27 at -60$^\circ$C, as shown in Fig.~\ref{fig:appstability}. Multiple measurements have been done and the uncertainty of this correction is determined to be 0.3\%. This combined temperature effect is corrected in the later analyses. Considering the changes of the refractive index of the liquid sample and the subsequent light spot change on the CCD, we use 0.4\% as the uncertainty of the measurement in the following studies.

\section{Solvents at low temperature}

Linear alkylbenzene is used by the Daya Bay and JUNO experiments as the LS solvent. It is a mixture of organic compounds containing a benzene ring and a linear alkyl chain of 10-13 carbons. LAB and LAB-based LS have a freezing point lower than -60$^\circ$C, suitable for a low temperature LS for TAO. We have measured the viscosity, density, specific heat capacity, and thermal conductivity of a LAB sample for TAO (the same LAB for the JUNO experiment) at different temperatures as shown in Table~\ref{tab:viscos}.

\begin{table*}[htb]
\begin{center}
\caption{Viscosity, density, specific heat capacity, and thermal conductivity of a typical LAB at different temperatures.\label{tab:viscos}}
\begin{tabular}{c c c c c }
\hline\hline
  Temperature & -20$^\circ$C &  -30$^\circ$C &  -40$^\circ$C &  -50$^\circ$C \\ \hline
  Viscosity (mm$^2$/s) & 54.2 & 114.7 & 283.4 & 802.5 \\ \hline
  Density (g/mL) & 0.896 & 0.902 & 0.908 & 0.914 \\ \hline
  Specific Heat Capacity (J/(g$\cdot$K)) & 1.784 & 1.761 & 1.740 & 1.727 \\ \hline
  Thermal Conductivity (W/(m$\cdot$K)) & 0.143 & 0.142 & 0.140 & 0.139 \\
\hline\hline
\end{tabular}
\end{center}
\end{table*}

Di-isopropylnaphthalene (DIN) and pseudocumene (PC) are also widely-used LS solvents. Especially, the LS made of DIN or PC has better pulse shape discrimination power than LAB-based LS. Adding 10\% DIN into LAB-based LS also improves the pulse shape discrimination significantly~\cite{Ko:2016owz}, which is attractive to reject fast neutron background for an experiment with shallow overburden like TAO. These two solvents are also studied here.

Although still in liquid state, LAB may turn cloudy at low temperature due to the water content in it. Usually the LAB exposed to air has a water content of tens ppm, varying at different humidities and temperatures, as well as the original water content in it. The saturated water content is $\sim200$~ppm at room temperature. Removing the trace amount of water in the solvent is thus critical to develop a low temperature LS. Bubbling the liquid with dry nitrogen is a safe and effective way to remove water without introducing impurities into LS. Table~\ref{tab:n2bubbling} shows the water content in a 100~mL sample when bubbling with 2~L/min dry nitrogen. An A831KF Coulometric moisture analyzer has been used to measure the water content in the samples. For all three solvents, the water content drops quickly in the first 30~minutes and reaches a plateau afterwards. Further improvements may require special processing to dry the nitrogen itself and improve the sealing of the nitrogen pipe and the vessel.

\begin{table}[htb]
\begin{center}
\caption{Water content (in ppm) in a 100 mL sample as a function of time when bubbling dry nitrogen with a flux of 2~L/min. \label{tab:n2bubbling}}
\begin{tabular}{lccccr}
\hline Time & 0~min & 5~min & 10~min & 30~min & 60~min\\
\hline LAB & 48.1 & 21.4 &9.8 & 5.1 & 4.4\\
 DIN & 40.2 & 18.3 & 10.3 & 5.5 & 4.6 \\
 PC & 245 & 168.5 & 113.3 & 105.1 & 102.4 \\
\hline
\end{tabular}
\end{center}
\end{table}

The transparency of the liquid samples at different temperatures are measured with the customized apparatus described in last section. As commonly used in UV-vis measurements, we define the {\it absorbance} $A_T$ of a sample at temperature $T$ as
\begin{equation}
 A_{T}=log_{10}\left(\frac{I_{0}\times{\varepsilon}_{T}}{I_{T}}\right) \,,
\end{equation}
where $I_{0}$ is the light intensity measured with an empty vessel in the light path, $\varepsilon_{T}$ is the combined temperature effect shown in Fig.~\ref{fig:appstability} as the red line, and $I_T$ is the measured light intensity of the sample. The {\it relative absorbance} $A_r$ is defined as
\begin{equation}
 A_{r}= A_T-A_{T_0} \,,
 \end{equation}
where $ A_{T_0}$ is the absorbance at the normalization temperature 20$^\circ$C.

\begin{figure}[htb]
\begin{center}
\includegraphics[width=7cm]{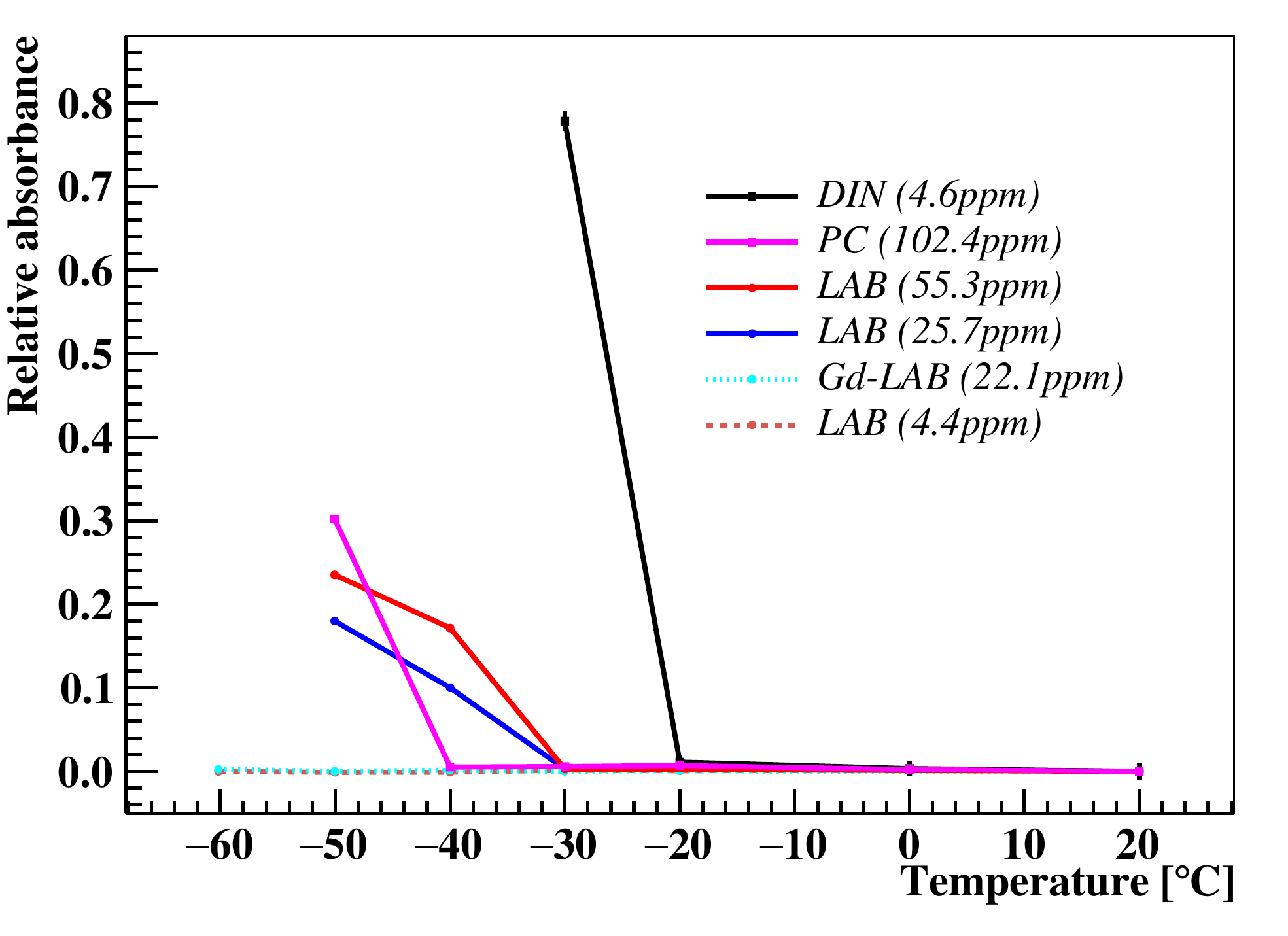}
\caption{Relative absorbance of DIN (4.6~ppm), PC (102.4~ppm), LAB (55.3~ppm), LAB (25.7~ppm), Gd-LAB (22.1~ppm), and LAB (4.4~ppm) at different temperatures, with the water content showing in parentheses. The Gd-LAB and LAB (4.4~ppm) curves are overlapping as neither of them has apparent transparency degradation at low temperatures. \label{fig:water_solvent} }
\end{center}
\end{figure}

The relative absorbance of a typical LAB samples with 55.3~ppm water starts to increase significantly when the temperature is lowered to -40$^\circ$C, as shown in Fig.~\ref{fig:water_solvent}. When reducing the water content to 25.7~ppm, the absorbance is still significant but reduces. For the LAB with 4.4~ppm water, there is no apparent transparency degradation from 20$^\circ$C down to -60$^\circ$C. Pure LAB will be used as the buffer liquid in the detector to shield the radioactivity from the outer tank, and to optically couple the acrylic vessel and the photosensors. With the studies here, we require the water content in the buffer LAB used in the TAO experiment should be less than 5~ppm. It could be reached by bubbling dry nitrogen for 30~min, but further studies needed to determine the parameters for large-scale processing.

Gadolinium-doped liquid scintillator (GdLS) will be used as the neutrino target and scintillator for TAO, similar to the Daya Bay experiment. The Gd-complex, a complex of Gd and a carboxylic acid, is dissolved into LAB to form the Gd-LAB solution, then mixed with the fluor 2,5-diphenyloxzzole (PPO) and the wavelength shifter p-bis-(o-methylstyryl)-benzene (bis-MSB) to form the GdLS in Daya Bay~\cite{Beriguete:2014gua}. The final GdLS for Daya Bay has 0.1\% Gd in mass fraction, 3~g/L PPO and 15~mg/L bis-MSB. The impact of the fluor and the wavelength shifter will be studied in next section. The transparency of Gd-LAB solution at low temperatures is shown in Fig.~\ref{fig:water_solvent}. With a similar nitrogen bubbling time of about 60~min, the Gd-LAB sample reaches a water content of $\sim 22$~ppm, comparing to the 4.4~ppm in LAB. The water content is high because water molecules participate the coordination in the Gd-complex, while the measurement of the Coulometric moisture analyzer includes both the free and the bound water. The Gd-LAB with 22~ppm water keeps good transparency down to -60$^\circ$C.

The DIN sample with 4.6 ppm water turns cloudy starting from -20$^\circ$C. We have also tested a sample with 10\% DIN in LAB, the transparency also degrades. PC has a freezing point of -43.78$^\circ$C. A PC sample with 102.4 ppm water shows large absorbance starting from -40$^\circ$C. DIN and PC are thus not suitable for the primary solvent for TAO. Adding a small fraction of DIN into LAB could be further evaluated to improve the particle identification, while adding PC may have safety difficulties due to its low flash point.

\begin{figure}[htb]
  \centering
  \includegraphics[width=3.5cm]{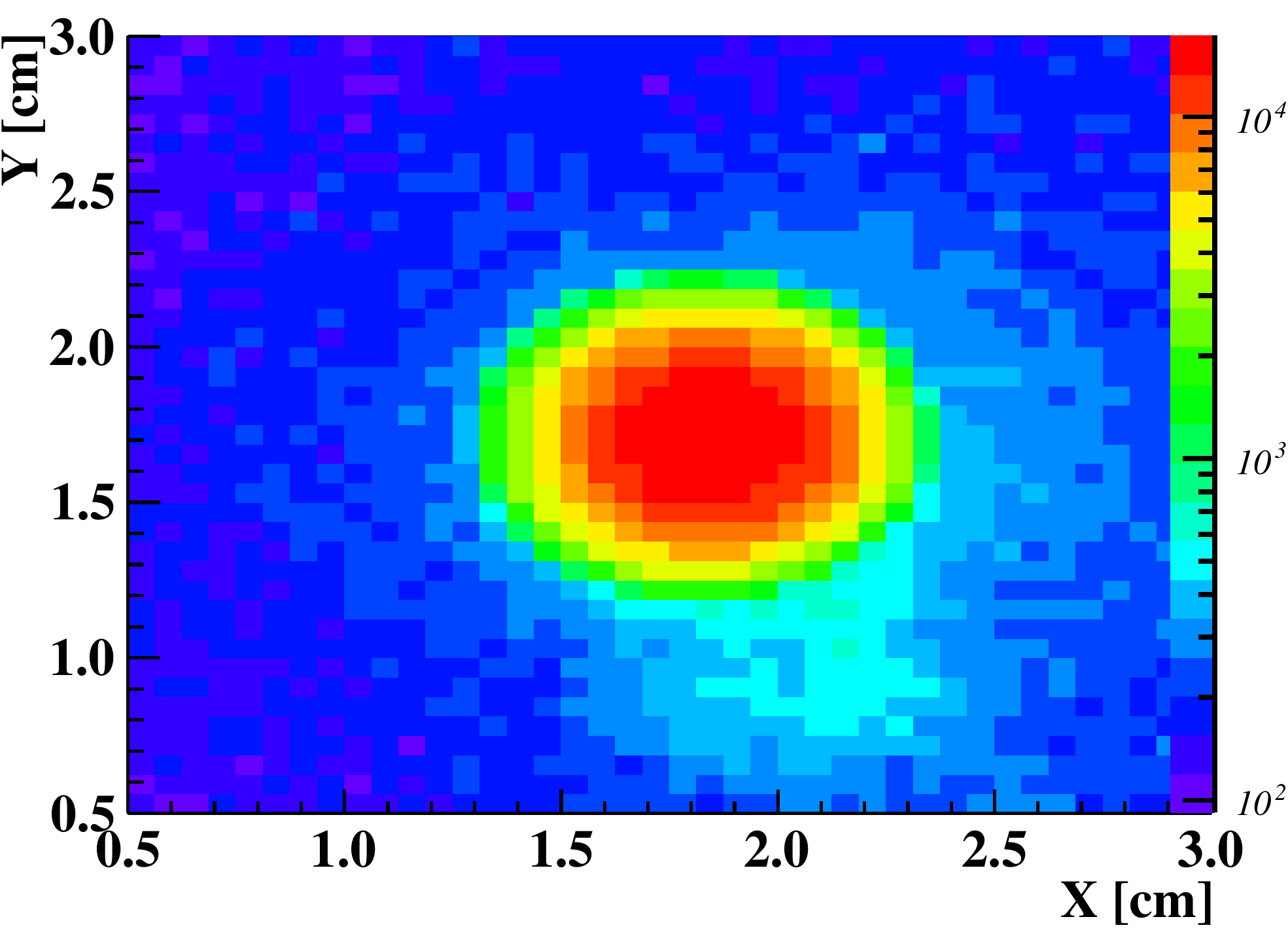}
  \includegraphics[width=3.5cm]{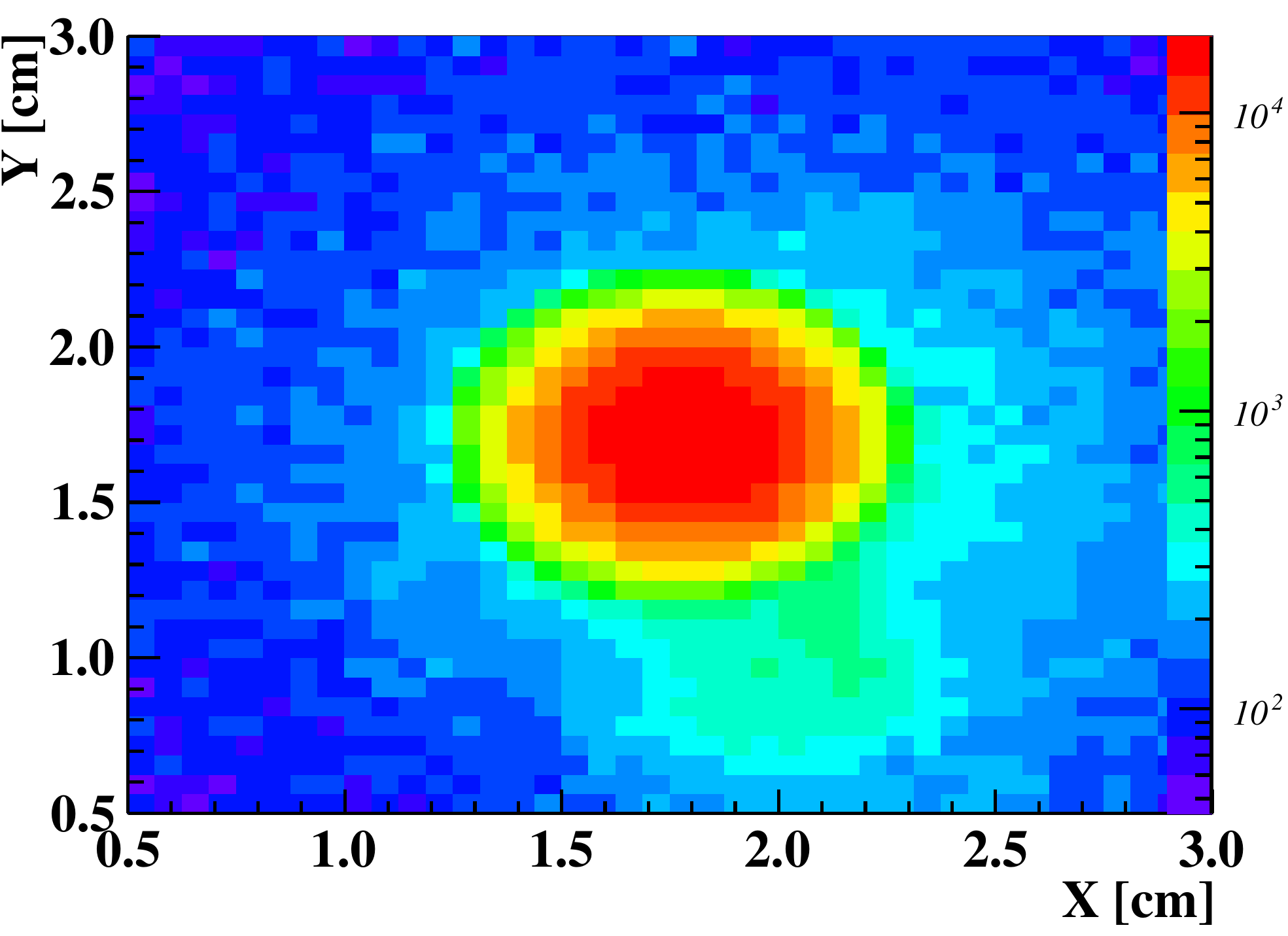}
  \includegraphics[width=3.5cm]{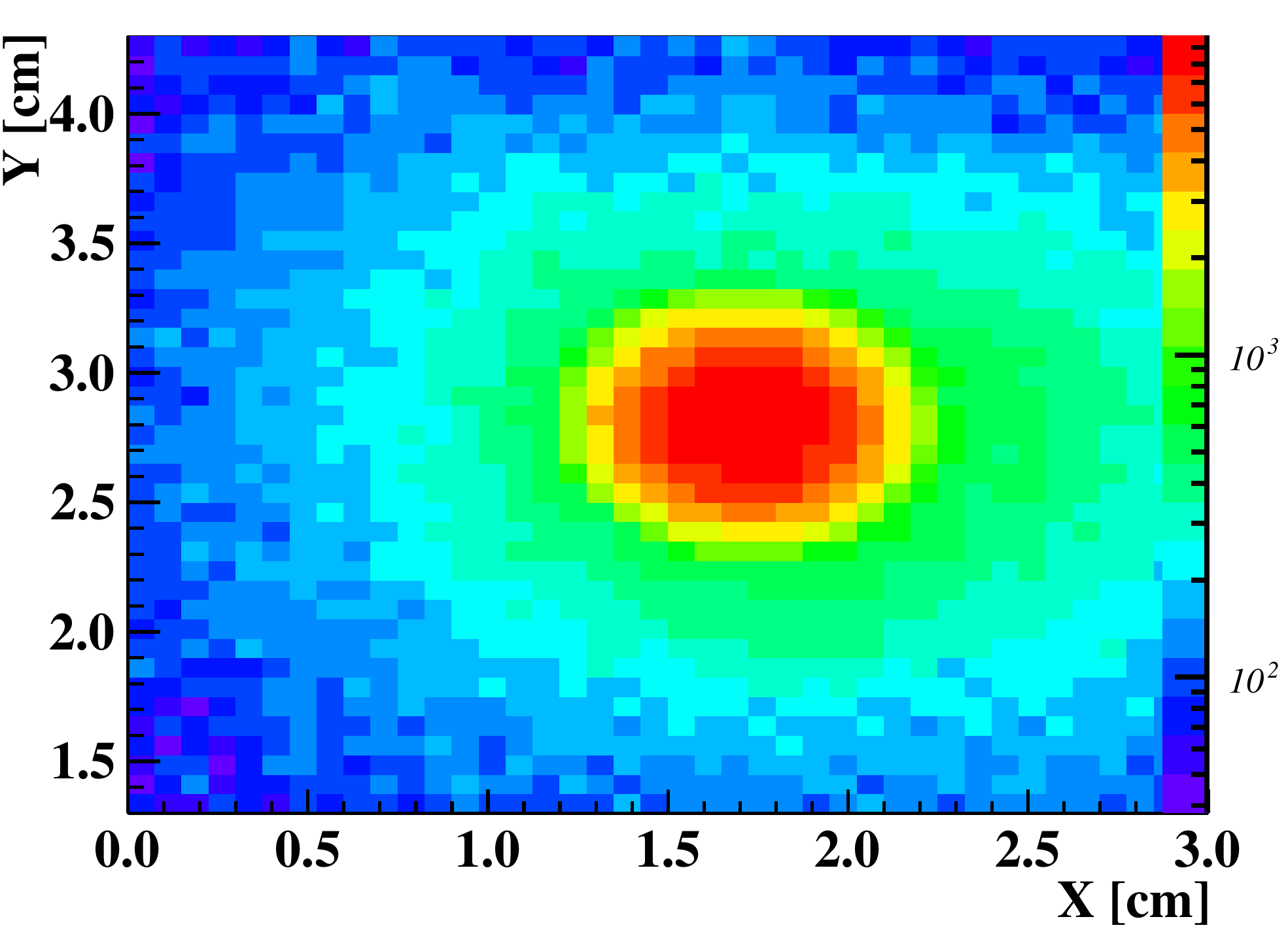}
  \includegraphics[width=3.5cm]{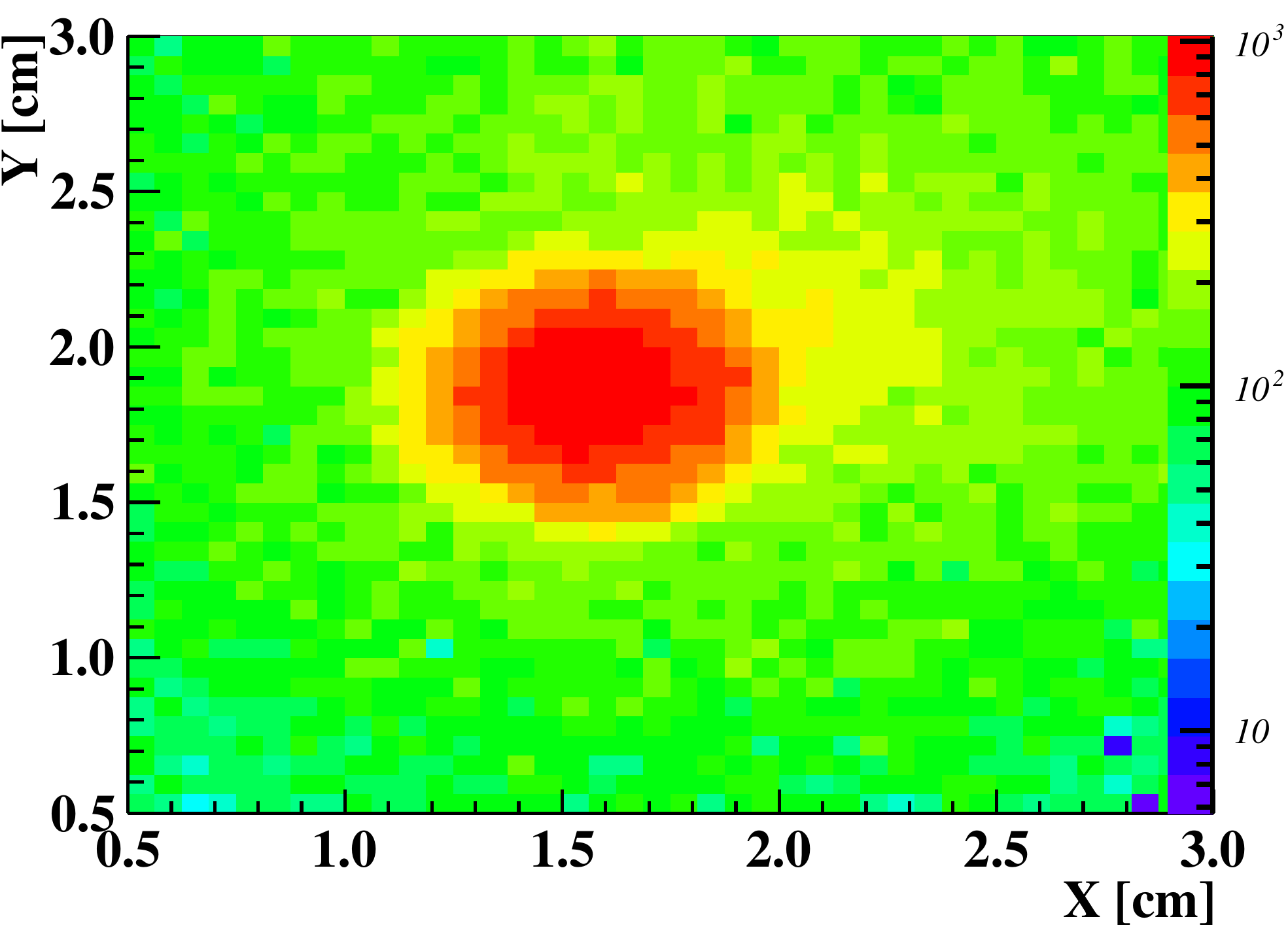}
\caption{Light spot pattern observed by the CCD. Top left: LAB (4.4 ppm water) at 20$^{\circ}$C; top right: LAB (4.4 ppm water) at -60$^{\circ}$C; bottom left: LAB (25.7 ppm water) at -50$^{\circ}$C; bottom right: DIN (4.6 ppm water) at -30$^{\circ}$C.
\label{fig:lightspot}}
\end{figure}

The absorbance measured above includes both light absorption and scattering. With a CCD as the photosensor, the light spot pattern on the CCD helps to distinguish these two effects and confirm the reliability of the measurements. As shown in Fig.~\ref{fig:lightspot}, the LAB sample with 4.4~ppm water has a clear spot on the CCD with sharp edge when measured at 20$^{\circ}$C. The light spot keeps almost unchanged but shows slightly fuzzy at -60$^\circ$C, while the absorbance, measured by integrating all CCD pixels, shows no apparent increase. For the LAB sample with 25.7 ppm water, the light spot extends a lot, which means the light scattering in the sample is large. For the DIN sample with 4.6~ppm water at -30$^\circ$C, the light scattering is prominent.

\section{Solubility of fluors at low temperature}

The Daya Bay GdLS recipe, with 0.1\% Gd by mass fraction, 3~g/L PPO, and 15~mg/L bis-MSB, shows large absorbance at low temperatures. We have demonstrated that Gd-LAB with 0.1\% Gd has good transparency at -50$^{\circ}$C or even -60$^{\circ}$C in last section. By changing the PPO and bis-MSB concentration in the LAB solution, we find that the large absorbance is due to precipitation of PPO and bis-MSB at low temperature.

\begin{figure}[htb]
\centering
\includegraphics[width=7cm]{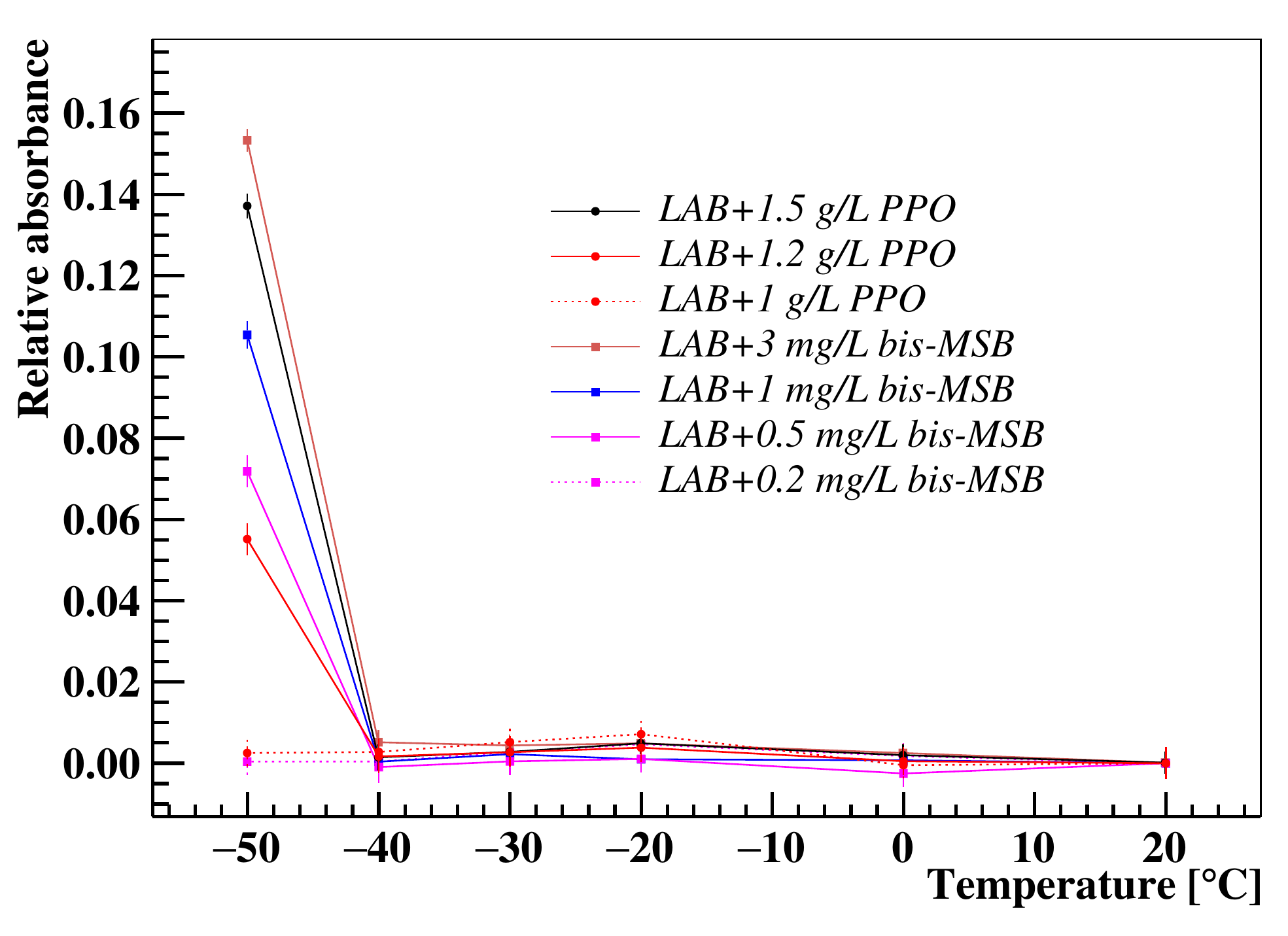}
\caption{Relative absorbance of solutions. The PPO and bis-MSB concentrations in LAB are showing in legend. \label{fig:fluor_solubility}}
\end{figure}

Fig.~\ref{fig:fluor_solubility} presents the relative absorbance of several samples at different temperatures, with different concentrations of PPO and bis-MSB in LAB. The measurements are done similar to that for the solvent studies in last section, and all samples have been bubbled with dry nitrogen for 30~min before the measurement. At \mbox{-50}$^{\circ}$C, the relative absorbance of the solution of 1~g/L PPO in LAB keeps the same as that at higher temperatures, while the relative absorbance of 1.2~g/L PPO in LAB increases apparently. Therefore, 
the solubility of PPO in LAB is between 1 and 1.2~g/L at \mbox{-50}$^{\circ}$C. Similarly, the solubility of bis-MSB in LAB is between 0.2 and 0.5~mg/L.

\begin{figure}[htb]
\begin{center}
  \includegraphics[width=7cm]{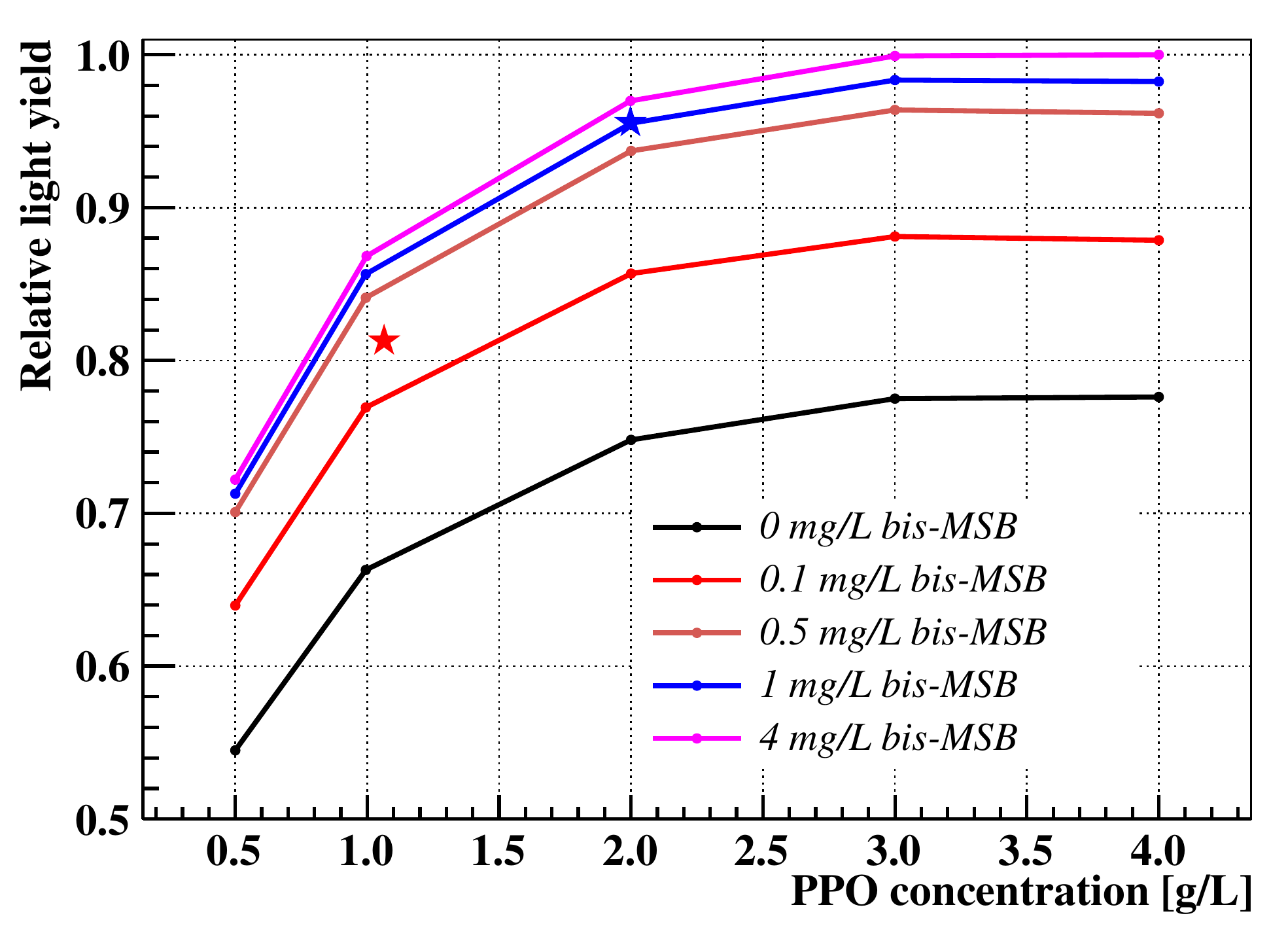}
  \caption{Simulation of the relative light yield of the LS with different concentrations of PPO and bis-MSB in LAB at room temperature. The red star corresponds to the LS with 1.2~g/L PPO and 0.5~mg/L bis-MSB, the highest possible concentration without co-solvent. The blue star corresponds to the LS with 2~g/L PPO and 1~mg/L bis-MSB, the allowed concentration if 0.05\% ethanol is added as co-solvent. \label{fig:lysim}}
\end{center}
\end{figure}

The light yield is critical for the TAO GdLS, since we anticipate an energy resolution as high as possible to study the fine structure in the reactor antineutrino spectrum. To predict the light yield of LS with different PPO and bis-MSB concentrations, Geant4~\cite{geant4} simulations are done with a simplified detector geometry and a newly-developed optical model~\cite{LSopticalModel}. The optical model takes into account the light emission and absorption processes by PPO and bis-MSB, and is tuned with the measurements in a Daya Bay detector with 12 different configurations on the PPO and bis-MSB concentrations. The predicted light yield curves are shown in Fig.~\ref{fig:lysim}. Even assuming the largest possible solubility of 1.2~g/L PPO and 0.5~mg/L bis-MSB in LAB at \mbox{-50}$^{\circ}$C, the relative light yield, shown as the red star in the plot, is only 82\% of the maximum, that with 4~g/L PPO and 4~mg/L bis-MSB.

To increase the light yield of LS at low temperature, a small fraction of ethanol is added into the LS to serve as co-solvent. The solubilities of PPO and bis-MSB in LAB is found to be improved significantly with a tiny amount of ethanol. However, adding ethanol, which has a flash point of 13$^\circ$C, increases the fire risk. The flash point is estimated to be $\sim40$$^\circ$C for a LS with 0.5\% ethanol by weight. To keep a reasonably high safety, we require the flash point to be $>100$$^\circ$C, corresponding to 0.05\% ethanol in the LS. With this amount of ethanol, the solubilities of the PPO and bis-MSB are found to be $>2$~g/L and $>1$~mg/L, respectively. The light yield of the LS would be 96\% of the maximum according to the simulation, as shown in Fig.~\ref{fig:lysim} as the blue star.

\begin{figure}[htb]
\begin{center}
  \includegraphics[width=7cm]{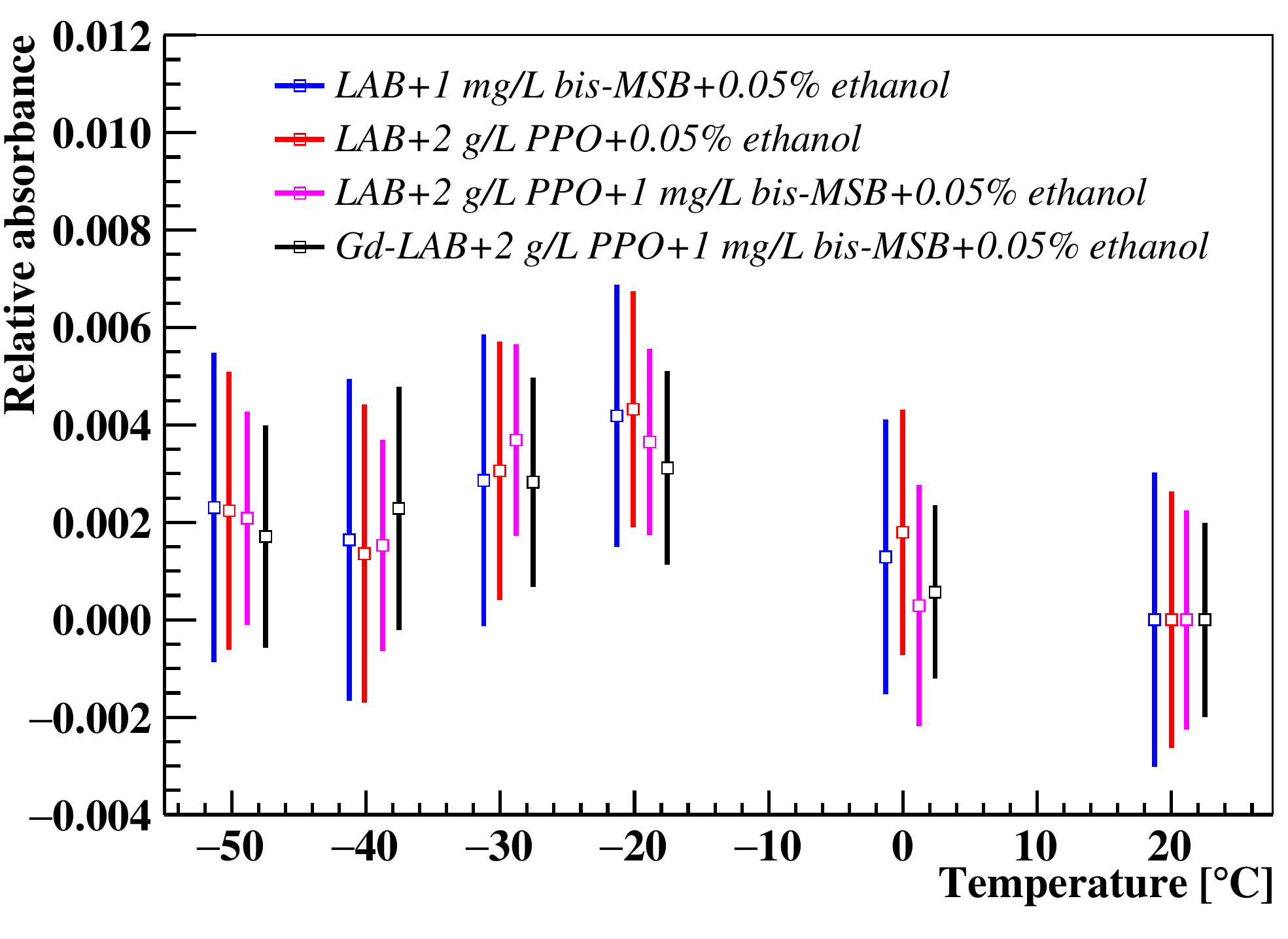}	
  \caption{Relative absorbance of four solutions with co-solvent added (0.05\% ethanol by weight) at different temperatures. The black data points correspond to the solution of Gd-LAB + 2~g/L PPO + 1~mg/L bis-MSB + 0.05\% ethanol, an excellent candidate GdLS to be used at low temperature. The data points are shifted slightly for visual clarity. \label{fig:lslowtemp}}
\end{center}
\end{figure}

The relative absorbances for samples with 0.05\% ethanol by weight are shown in Fig.~\ref{fig:lslowtemp}. The four samples, LAB with 1~mg/L bis-MSB, LAB with 2~g/L PPO, LAB with 2~g/L PPO and 1~mg/L bis-MSB, and Gd-LAB with 2~g/L PPO and 1~mg/L bis-MSB, all show good transparency down to -50$^\circ$C.

\section{Light yield of the new LS}

Adding co-solvent may have an impact on the light yield. A preliminary study of the light yield of a LS with 2~g/L PPO, 1~mg/L bis-MSB, and 0.05\% ethanol in LAB, which has the same recipe as the preferred TAO GdLS but without Gd, has been done and compared to the JUNO LS, with 2.5~g/L PPO and 3~mg/L bis-MSB in LAB. The measurement is done with a customized apparatus, in which six PMTs are coupled to a cubic acrylic vessel of 5-cm dimension. Scintillation light is excited by the internal conversion electron of a $^{207}$Bi source. The relative light yield of the new LS is about 96\% of the JUNO LS at room temperature, consistent with the simulation. Therefore, adding co-solvent and Gd-complex into the LS has no apparent impact to the light yield.

All the light yield measurements and simulations above are done at room temperature. At low temperature,
the light yield usually increases due to less thermal quenching. At the same time, the nonlinear energy response and the particle-dependent energy quenching effect are also important for the TAO experiment. They will be studied later in details.

\section{Conclusion}

We have developed a recipe of liquid scintillator that works at low temperatures down to -50$^\circ$C. The solubilities of the fluor PPO and the wavelength shifter bis-MSB in the solvent LAB reduce significantly to $<1.2$~g/L PPO and $<0.5$~mg/L bis-MSB at -50$^\circ$C. By adding 0.05\% ethanol as co-solvent, their solubilities raise to $>2$~g/L PPO and $>1$~mg/L bis-MSB. Usually the liquid scintillator based on LAB has a water content of tens ppm, leading to lower transparency at low temperature. By bubbling dry nitrogen to remove the water, the LS transparency keeps unchanged at -50$^\circ$C comparing to that at room temperature. Gadolinium of 0.1\% by weight is doped into the LS as Daya Bay, by synthesizing a complex of Gd and a carboxylic acid.
Finally, a recipe of GdLS with 0.1\% Gd by weight, 2~g/L PPO, 1~mg/L bis-MSB, and 0.05\% ethanol by weight in the solvent LAB has been synthesized and tested. The new GdLS shows good transparency at -50$^\circ$C and also good light yield.

\section{Acknowledgement}

This work is supported by the National Nature Science Foundation of China (Grant No.11775252 and 11225525) and the Strategic Priority Research Program of Chinese Academy of Sciences (XDA10010900).

\bibliographystyle{unsrt}
\bibliography{lowtempLS}

\begin{thebibliography}{10}

\bibitem{Cowan1956}
C.~L. Cowan, F.~Reines, F.~B. Harrison, H.~W. Kruse, and A.~D. McGuire.
\newblock {Detection of the free neutrino: A Confirmation}.
\newblock {\em Science}, 124:103--104, 1956.

\bibitem{kamland2011}
A.~Gando et~al.
\newblock {Constraints on $\theta_{13}$ from A Three-Flavor Oscillation
  Analysis of Reactor Antineutrinos at KamLAND}.
\newblock {\em Phys. Rev. D}, 83:052002, 2011.

\bibitem{borexino2008}
G.~Alimonti et~al.
\newblock {The Borexino detector at the Laboratori Nazionali del Gran Sasso}.
\newblock {\em Nucl. Instrum. Meth. A}, 600:568--593, 2009.

\bibitem{An:2015jdp}
Fengpeng An et~al.
\newblock {Neutrino Physics with JUNO}.
\newblock {\em J. Phys.}, G43(3):030401, 2016.

\bibitem{An:2015nua}
F.~P. An et~al.
\newblock {Measurement of the Reactor Antineutrino Flux and Spectrum at Daya
  Bay}.
\newblock {\em Phys. Rev. Lett.}, 116(6):061801, 2016.
\newblock [Erratum: Phys. Rev. Lett. 118, no.9, 099902(2017)].

\bibitem{Abe:2014bwa}
Y.~Abe et~al.
\newblock {Improved measurements of the neutrino mixing angle $\theta_{13}$
  with the Double Chooz detector}.
\newblock {\em JHEP}, 10:086, 2014.
\newblock [Erratum: JHEP 02, 074 (2015)].

\bibitem{Ko:2016owz}
Y.J. Ko et~al.
\newblock {Sterile Neutrino Search at the NEOS Experiment}.
\newblock {\em Phys. Rev. Lett.}, 118(12):121802, 2017.

\bibitem{Bak:2018ydk}
G.~Bak et~al.
\newblock {Measurement of Reactor Antineutrino Oscillation Amplitude and
  Frequency at RENO}.
\newblock {\em Phys. Rev. Lett.}, 121(20):201801, 2018.

\bibitem{Huber:2011wv}
Patrick Huber.
\newblock {On the determination of anti-neutrino spectra from nuclear
  reactors}.
\newblock {\em Phys. Rev. C}, 84:024617, 2011.
\newblock [Erratum: Phys.Rev.C 85, 029901 (2012)].

\bibitem{Mueller:2011nm}
Th.A. Mueller et~al.
\newblock {Improved Predictions of Reactor Antineutrino Spectra}.
\newblock {\em Phys. Rev. C}, 83:054615, 2011.

\bibitem{Estienne:2019ujo}
M.~Estienne et~al.
\newblock {Updated Summation Model: An Improved Agreement with the Daya Bay
  Antineutrino Fluxes}.
\newblock {\em Phys. Rev. Lett.}, 123(2):022502, 2019.

\bibitem{junotaocdr}
Angel Abusleme et~al.
\newblock {TAO Conceptual Design Report}, 2020.
\newblock arXiv:2005.08745.

\bibitem{Ziegler1956}
C.A. Ziegler, H.H. Seliger, and I.~Jaffe.
\newblock Three ways to increase efficiency of liquid scintillators.
\newblock {\em Nucleonics}, 14(5):84, 1956.

\bibitem{HOMMA198791}
Yoshio Homma, Yuko Murase, and Kazue Sonehara.
\newblock The effect of temperature on fluorescence for liquid scintillators
  and their solvents.
\newblock {\em Int. J. Rad. Appl. Instrum. A}, 38(2):91 -- 96, 1987.

\bibitem{Xia:2014cca}
Dong-Mei Xia et~al.
\newblock Temperature dependence of the light yield of the lab-based and
  mesitylene-based liquid scintillators.
\newblock {\em Chin.Phys.C}, 38(11):116001, 2014.

\bibitem{Sorensen:2018skx}
A.~Sörensen, S.~Hans, A.R. Junghans, B.v. Krosigk, T.~Kögler, V.~Lozza,
  A.~Wagner, M.~Yeh, and K.~Zuber.
\newblock {Temperature quenching in LAB based liquid scintillator}.
\newblock {\em Eur. Phys. J. C}, 78(1):9, 2018.

\bibitem{Beriguete:2014gua}
Wanda Beriguete et~al.
\newblock Production of a gadolinium-loaded liquid scintillator for the daya
  bay reactor neutrino experiment.
\newblock {\em Nucl.Instrum.Meth.A}, 763:82--88, 2014.

\bibitem{Djurcic:2015vqa}
Zelimir Djurcic et~al.
\newblock {JUNO Conceptual Design Report}, 2015.
\newblock arXiv:1508.07166.

\bibitem{geant4}
S.~Agostinelli et~al.
\newblock {GEANT4: A Simulation toolkit}.
\newblock {\em Nucl. Instrum. Meth.}, A506:250--303, 2003.

\bibitem{LSopticalModel}
Yan Zhang, Ze-Yuan Yu, Xin-Ying Li, Zi-Yan Deng, and Liang-Jian Wen.
\newblock {A complete optical model for liquid-scintillator detectors}.
\newblock {\em Nucl. Instrum. Meth. A}, 967:163860, 2020.

\end{thebibliography}

\end{document}